\documentclass[12pt]{article} 
\usepackage{epsfig}
\usepackage{graphicx,amsmath}  

\def\fun#1#2{\lower3.6pt\vbox{\baselineskip0pt\lineskip.9pt
  \ialign{$\mathsurround=0pt#1\hfil##\hfil$\crcr#2\crcr\sim\crcr}}}

\newcommand{\be}{\begin{equation}}
\newcommand{\ee}{\end{equation}}
\newcommand{\ba}{\begin{eqnarray}}
\newcommand{\ea}{\end{eqnarray}}
\newcommand{\bg}{\begin{gather}}
\newcommand{\foma}{\end{gather}}

\title{ Two comments to utilization of structure function approach in DIS experiments.
}

\author{
E.~Kuraev$^a$, M.~Galynskii$^b$, A.~Ilyichev$^c$
\vspace{4mm}
\\
\small\sl $^a$ Joint Institute for Nuclear Research, 141980 Dubna, Russia\\
\small\sl $^b$ Institute of Physics NAS, 220072 Minsk, Belarus \\
\small\sl $^c$ National Centre of Particle and High Energy Physics , 220040 Minsk,  Belarus \\
}

\date{}
\begin{document}
\maketitle

\begin{abstract}
The "returning to resonance" mechanism can
be used to obtain the simple procedure of taking RC to DIS cross section into account,
in frames of Drell-Yan picture. Iteration procedure is proposed.

Kinematic region $y \to 1$ can be described in frames of Drell-Yan picture using structure
function approach. The large RC in the lowest order reflect Sudakov form factor suppression,
which can be taken into account in all orders of perturbation theory.
Basing on explicit calculation in two lowest orders of perturbation theory we construct the
cross section in $y \to 1$ region obeying renormalization group equations and include the Sudakov-like
form factor suppression.
\end{abstract}
\newpage
\section*{Comment 1}
It's now accepted the DIS cross section with taken into account the radiative corrections (RC)
have form of a cross section of Drell-Yan process (see \cite{KFM} and references therein):
\begin{gather}
 \frac{d^2\sigma(p_1,p_2)}{dQ^2 dy}=\int\limits ^1_{z_{1m}} dz_1 \int \limits ^1_{z_{2m}}dz_2 \,\frac{1}{z_2^2} {\cal{D}}(z_1,t)
{\cal{D}}(z_2,t) \frac{d^2 \tilde{\sigma}^{hard}(z_1 p_1,p_2/z_2)} {d\tilde{Q}^2 d\tilde{y} }
\left( 1+\frac{\alpha}{\pi} K \right),\\
t=\frac{\alpha}{\pi}L,\quad L=\ln(Q^2/m_e^2),\quad p_1^2=p_2^2 =m_e^2, \quad Q^2=-(p_1 -p_2)^2 \gg m_e^2, \\
 y=\frac{2p_1 q}{2p_1 P}, \quad q=p_1-p_2,
\nonumber
\end{gather}
with scaling parameters of the hard cross section:
\begin{eqnarray}
\tilde{Q}^2&=&\frac{z_1}{z_2} Q^2 \, , \, \tilde{y}=1- \frac{1-y}{z_1 z_2}\,, \,\nonumber
z_{1m}=\frac{1+z_{th}-y}{1-xy}\, , \\
 z_{2m}&=&\frac{1-y+xyz_1}{z_1-z_{th}}\, , \, z_{th}=\frac{2 m_{\pi} M}{2p_1 P} \ll 1\, , \, P^2=M^2. \nonumber
\end{eqnarray}

Performing the integration in equation (1) over partonic energy fraction $z_{1}$ it is necessary to
keep in mind two enhancing tendencies. The main is related with the rapid decreasing of the hard cross
section with increasing of $Q^2$:
\ba
\frac{d \tilde{\sigma}^{hard}}{d Q^2 \d y}&=&
\biggl(\frac{1}{1-\Pi(Q^2)}\biggr)^2 \frac{d \sigma^{hard}_B}{d Q^2 d y},
\ea
where
\ba
\frac{d\sigma^{hard}_B}{d Q^2 d y}&=&\frac{4\pi\alpha^2(Q^2)}{Q^4
y}[(1-y-x^2y^2\frac{M^2}{Q^2})F_2(x,Q^2)+xy^2F_1(x,Q^2)],
\ea
and $\Pi(Q^2)$ is the polarization operator of virtual photon.
The explicit value of $K$-factor can be found in paper of A. Afanasev et al. [1].

Another one is related with rather slow tendency of structure functions:
\ba
{\cal{D}}(z_1,L)\approx \varepsilon (1-z_1)^{-1+\varepsilon}\approx\delta(1-z_1),\quad \varepsilon=\frac{2\alpha}{\pi}L\,.
\ea
These tendencies are struggling. For the aim to extract the main contribution arising from the first
mechanism (known in colliding beams experiments as a "returning to resonance" one).

Let us put the right hand side (r.h.s.) of (1) in form:
\ba
\int \limits _{z_{1m}}^1\, \frac{d z_1}{z_1^2}\, \Psi(z_1)=\left(\frac{1}{z_{1m}}-1\right)\Psi(z_{1m})+\int \limits _{z_{1m}}^1
\frac{dz_1}{z_1}(1-z_1)\frac{d}{dz_1}\Psi(z_1)\;.
\ea
The second term in r.h.s. of (4) is substantially smaller than the first one at $z_{1m} \ll1$. The enhancement
tendencies in the second term integrand now become of the same rate: $\varepsilon z_1^{-1}(1-z_1)^{-1+\varepsilon}$.

During the integration over $z_2$ (which is dropped here) only one enhancement tendency caused by the
structure function ${\cal D}(z_2,L)\sim \delta(1-z_2)$ is appeared. We will not touch it here. The
expression (6) provides the application of iteration procedure: using subsequently $\Psi$ as an
experimental data.

\section*{Comment 2}
It's widely believed that
due to huge RC which exceed lowest order more then 100\% \cite{Bardin}
the kinematic region $1-y \ll 1$ in DIS experiments cannot be described correctly. This fact is the reason why the experimental results at $y > 0.8$ region as a rule are excluded from data processing.
However we argue that for the correct description of high $y$ region RC
all orders of perturbation theory (PT) must be taken into account.

For the solution of this task the renormalization group approach
is modified in such a way to include Sudakov-type suppression formfactor.
We will consider here the
experimental set-up with no emission of hard photons along an initial lepton.


For this aim let us consider two lowest order RC of PT. The
emission of additional soft pions and soft pairs of the same order
of energy as the one $\varepsilon_2$ of a scattered lepton do not
exceeding $\Delta\varepsilon \ll \varepsilon$ becomes relevant:
\ba \Delta \varepsilon\sim \varepsilon_2=\varepsilon (1-y)\ll
\varepsilon_1=\varepsilon. \ea The cross section within RC can be
put in form : \ba
\frac{d\sigma}{d\sigma_B}=1+\delta,\;\delta=\frac{\alpha}{\pi}\Delta^{(1)}+\left(\frac{\alpha}{\pi}\right)^2\Delta^{(2)}+...,
\ea while the lowest order RC are \ba
\Delta^{(1)}=(l_t-1)\left(\ln\frac{\Delta\varepsilon}{\varepsilon_1}+\ln\frac{\Delta\varepsilon}{\varepsilon_2}\right)+
\frac{3}{2}l_t-\frac{1}{2}\ln^2(1-y) -\frac{\pi^2}{6}-2+{\rm
Li}_2\left(\frac{1+c}{2}\right), \ea with \ba
-t=2\varepsilon^2(1-y)(1-c)\gg m_e^2,\;
l_t=\ln\left(\frac{-t}{m_e^2}\right), \; c=\cos \theta\;, \ea
where\ $\theta=\widehat{\vec{p}_1\vec{p}_2}$ and $\varepsilon_2$
are the scattering angle and the energy of the scattered lepton in
the laboratory frame. The mentioned above reasons allow us to put:
\ba
\ln\frac{\Delta\varepsilon}{\varepsilon_1}+\ln\frac{\Delta\varepsilon}{\varepsilon_2}=\ln(1-y).
\ea As a result we have some deviation from the well-known
$\Delta$-part of evolution equation kernel \ba
P^{(1)}(x)=\lim_{\Delta\to
0}\biggl\{\delta(1-x)P^{(1)}_{\Delta}+\theta(1-x-\Delta)\frac{1+x^2}{1-x}\biggr\}.
\ea Here in the our approach $\theta$-part does not work,
$\Delta=1-y$, (see the term containing $l_t$ in (9)): \ba
P_\Delta^{(1)}=2\ln\Delta+\frac{3}{2}\to
\left(2\ln(1-y)+\frac{3}{2}\right)-\ln(1-y). \ea

At the second order of PT the emission of two soft photons and soft pair (with total energy not
exceeding
$\Delta\varepsilon$) as well as a single photon emission with 1-loop RC and, finally the 2-loop
virtual corrections must be taken into account:
$\Delta^{(2)}=\delta_{\gamma\gamma}+\delta_{sp}$. We will not consider here the contribution from emission of real and
virtual pairs. It can be taken into account by replacing the coupling constant by the moving one.

Contributions to RC from virtual and real photons emission have a form:
\ba
\delta_{\gamma\gamma}=\frac{1}{2}(\Delta^{(1)})^2-\frac{\pi^2}{3}(l_t-1)^2+\frac{3}{2}l_t\left(2+\frac{\pi^2}{6}-
{\rm Li }_2\left(\frac{1+c}{2}\right)\right)+O(1).
\ea
This result agrees with renormalization group (RG) predictions \cite{KF} at $y=0$ and,
in addition, contains the terms
of type $\ln^2(1-y), \, l_t\ln(1-y)$, which become relevant in the limit $y \to 1$.

Let us discuss this points more closely. We suppose no hard photon
emission by the initial lepton which can provide the "returning to
resonance" mechanism. Really this mechanism for the case
$\varepsilon_2/\varepsilon=1-y<<1$ will correspond to very small
transversal momentum squared $Q_1^2\sim \epsilon^2 (1-y)^2<< Q^2$.

Let us now average DIS cross section over small interval $\tilde{Q}^2\sim Q^2$ introducing the
additional integration in the right part of formula (1):
\ba
\int d \tilde{Q}^2 \delta((z_1xQ^2/z_2)-\tilde{Q}^2),\; x=1-(\Delta \epsilon/\epsilon).
\ea
Small variations of transfer momentum arises from RC-emission of soft real and virtual partons
(photons and leptons). Using the flatness of the hard cross section in this region we obtain for the ratio
of DIS cross sections with and without RC:

\ba
\frac{d\sigma}{d\sigma_B}=F(x,t)=\int \int D(z_1,t)D(z_2,t)d z_1d z_2 \theta(xz_1-z_2).
\ea
Using the differential evolution equations for nonsinglet structure functions $D(x,t)$
\ba
\nonumber
\frac{\partial D}{\partial t}=\frac{\alpha(t)}{2\pi}\int \limits _x^1\frac{d y}{y}P \left( \frac{x}{y} \right) D(y,t),\;\; D(y,0)=\delta(1-y).
\ea
One can obtain a differential equation for $F$
\ba
\frac{\partial F}{\partial t}=\frac{\alpha(t)}{\pi}\int \limits _x^1 d z P\left( \frac{x}{z}\right) F(z,t), \;\;F(x,0)=1.
\ea
This equation was solved in paper [2]:
\ba
F(x,t)=\left (\ln\frac{1}{x}\right)^{2\chi} \frac{\exp [\chi(3/2-2C_E)]}{\Gamma(1+2\chi)}
,\;\; \chi=-3 \ln (1-\frac{1}{3}t).
\ea

Terms containing $\ln(1-y)$ are not taken into account in
evolution procedure. We argue here that there is a reason to take
them into account as a general factor which can be obtained from
the known factor of Yienie, Frautchi and Suura \cite{YFS}, with
replacement of logarithm of ratio of photon mass to lepton mass by
$\ln \Delta , \Delta=\Delta\varepsilon/\varepsilon$ with
accordance with Bloch-Nordsick theorem.

Replacing $\ln(1/x)=1-y$ we obtain for DIS cross section:
\ba
&&  \frac{d\sigma}{d\sigma _B}\biggl|_{y\to 1}=R(1+\frac{\alpha}{\pi}K), \quad
\\ \nonumber
&&R= \frac{1}{(1-\Pi(Q^2))^2}
\frac{(1-y)^{2\chi}}{\Gamma(1+2\chi)} \exp\left((3/2-2C_E)\chi
-\frac{\alpha}{2\pi}(\ln^2(1-y)+2l_t\ln(1-y))\right), \\ \nonumber
&&|K|\sim 1, \quad
\chi=-3\ln\left(1-\frac{\alpha}{3\pi}l_t\right)=
\frac{\alpha}{\pi}l_t+\frac{\alpha^2}{6\pi^2}l_t^2+\dots . \ea
with $C_E=0.577$ is Euler constant and $d\sigma_B$ is the DIS
cross-section in Born approximation. One can be convinced in the
agreement (19) with the results of lowest order calculation
(13,14) up to non-leading terms, which are parameterized in form
of $K$-factor. The formula (19) provides us for $|K|\sim 1$ with
the accuracy on the level of $1\%$. The behaviour of quantity
$R(x,y)$ for different values of Bjorken parameter $x$ at HERMES
kinematic conditions is illustrated in fig.1.

\section{Acknowledgment}
One of us (E.K.) is grateful to Heisenberg-Landau grant 2001-02 and  to SCAR JINR.
 Two of us (E.K. and A.I.) are grateful to DESY  HERMES staff for hospitality and to  V.S. Fadin for discussion.

\hspace{1.0cm}
\begin{figure}[ht]

\vspace{-4.0cm}
\centerline{
\leavevmode \epsfxsize=0.70\textwidth \epsfbox{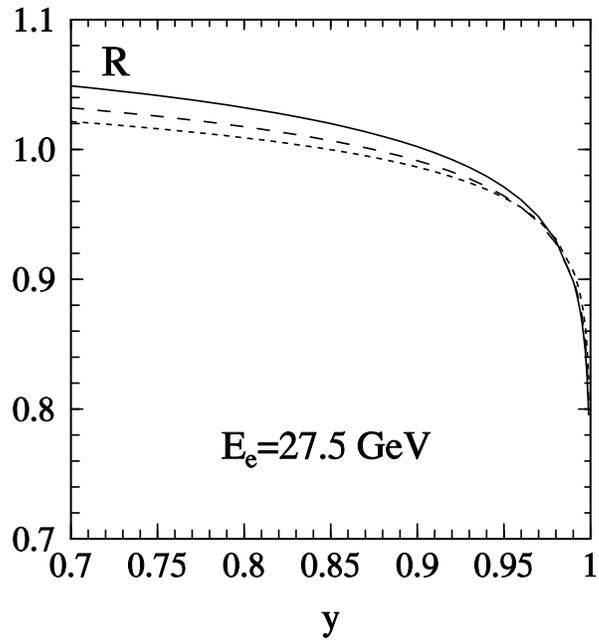} }
\vspace{-1.50cm} \nonumber \caption{\it The ration
$R=d\sigma/d\sigma_B$ for $x=0.1$ (solid line), $x=0.01$ (dashed
line) and $x=0.001$ (dotted line) versus $y$ for $0.7\leq y \leq
0.999$ at HERMES kinematic condition. } \label{fig1}
\end{figure}

\end{document}